\documentclass{article}

\usepackage{emulateapj}
\usepackage{onecolfloat}
\usepackage{graphicx} 
\usepackage{fancyheadings} 
\usepackage{ulem}
\usepackage{rotating}
\usepackage{lscape}

\def\lya{Ly$\alpha$ }

\def\kms{km~s$^{-1}$ }
\def\cm#1{\, {\rm cm^{#1}}}
\def\N#1{{N({\rm #1})}}

\def\sci#1{{\rm \; \times \; 10^{#1}}}
\def\ltk{\left [ \,}

\def\rtk{\, \right  ] }

\def\mkms{{\rm \; km\;s^{-1}}}
\newcommand{\tskip}{\tablevspace{3pt}}

\begin{document}

\twocolumn[%
\submitted{ApJL: Accepted November 11, 1998}

\title{THE PHYSICAL NATURE OF THE LYMAN LIMIT SYSTEMS}

\author{ JASON X. PROCHASKA\altaffilmark{1} \\
Department of Physics, and Center for Astrophysics and Space Sciences; \\
University of California, San Diego; \\
C--0424; La Jolla; CA 92093\\}

\begin{abstract} 

We analyze Keck HIRES observations of a Lyman limit system at $z=2.652$
toward Q2231$-$00.  These observations afford the most comprehensive
study of the physical properties of a LL system to date.
By comparing the ionic column densities for 
Fe$^+$, Fe$^{++}$, Si$^+$, and Si$^{3+}$ against calculations derived
from the CLOUDY software package, we have strictly constrained
the ionization state of this system.  This has enabled us to calculate
accurate abundances of a Lyman limit system for the first time at $z > 2$, 
e.g., $\lbrack$Fe/H$\rbrack$ = $-0.5 \pm 0.1$.  
We also derive a total hydrogen column density, $\log \N{H} = 20.73 \pm 0.2$, 
which is comparable to values observed for the damped \lya systems.

The system is special for exhibiting CII*~1335 absorption, allowing
an estimate of the electron density, $n_e = 6.5 \pm 1.3 \sci{-2} \, \cm{-3}$.
Coupling this measurement with our knowledge of the ionization state,
we derive the following physical properties: (1) hydrogen volume density,
$n_H = 5.9 \pm 1.2 \sci{-2} \, \cm{-3}$, (2) path length, 
$\ell = 3 \pm 1.6$~kpc, and (3) ionizing intensity, 
$\log J_{912} = -20.22 \pm 0.21$.
We point out that a number of the physical properties 
(e.g.\ $\lbrack$Fe/H$\rbrack$, $\N{H}$,
$n_H$) resemble those observed for the damped \lya systems, which 
suggests this system may be the photoionized analog of a damped system.
The techniques introduced in this Letter should be applicable to a 
number of Lyman limit systems and therefore enable a survey of their
chemical abundances and other physical properties.
\end{abstract}

\keywords{quasars:individual (Q2231$-$00) --- absorption lines ---
galaxies: abundances}]

\altaffiltext{1}{Visiting Astronomer, W.M. Keck Telescope.
The Keck Observatory is a joint facility of the University
of California and the California Institute of Technology.}

\pagestyle{fancyplain}
\lhead[\fancyplain{}{\thepage}]{\fancyplain{}{PROCHASKA}}
\rhead[\fancyplain{}{THE PHYSICAL NATURE OF THE LYMAN LIMIT SYSTEMS}]{\fancyplain{}{\thepage}}
\setlength{\headrulewidth=0pt}
\cfoot{}

\section{INTRODUCTION}

Observations of the Quasar Absorption
Line (QAL) systems provide an efficient means for probing the
physical conditions of the early universe.
Intermediate resolution surveys have enabled accurate measurements
on the evolution of the number density of the QAL systems, including
the damped \lya systems (\cite{wol86,wol95}), the Lyman limit
(LL) systems (\cite{sar89}), and the \lya forest clouds 
(\cite{srgt80}).  Until recently, the absence of a comprehensive
physical description of the various QAL systems has limited the 
impact of these surveys.
With the wealth of data afforded by 
echelle spectrographs on 10m class telescopes,
one is now capable of making direct measurements
of the physical properties of individual QAL systems and, in
turn, the role of these systems in galaxy
formation and the evolution of the early universe.

Over the past few years, observations
of the damped \lya systems -- QAL systems with neutral hydrogen column
densities $\N{HI} > 2 \sci{20} \cm{-2}$ --
have provided detailed measurements on their
chemical composition, ionization state, dust content, 
metallicity, and kinematic characteristics (\cite{ptt94,pro96,lu96b,pro97a}). 
These observations present vital clues to galaxy formation theory.
Thus far, little research has focused on the physical properties of the
Lyman limit systems, QAL systems optically thick at the
Lyman continuum.  Owing to the
challenges associated with accurately
determining the ionization state of these
partially ionized systems,  their metallicity and chemical abundances
have been poorly constrained.  
Steidel (1990) performed a survey of the physical properties of
$z \approx 3$ LL systems, but the majority of his observations
resulted only in limits to the abundances and other properties.
With Keck, thus far researchers have focused 
on the kinematic characteristics 
of the MgII systems (\cite{chrl96}), the majority of which
are LL systems.  
Their efforts have provided keen insight into these systems at
moderate redshift $z \approx 1$, but have been somewhat limited by
difficulties in identifying their ionization state.

In this Letter, we present techniques aimed at 
determining the physical characteristics of a single LL system,
including measurements of the ionization state, chemical abundances,
electron density, and the intensity of ionizing radiation.
While the results from this single LL system are
not necessarily representative of all LL systems, the techniques outlined
in this letter will be applicable to further LL studies, particularly
when $\N{HI} > 10^{18} \cm{-2}$.
By applying these methods to a significant number of LL systems
at high redshift we will investigate their chemical history,
contrast their properties with those of the damped \lya systems,
and thereby probe physical conditions in the early universe.
Section 2 presents the observations and measurements of the 
ionic column densities and the ionization state.  In $\S$3,
we determine elemental abundances by
making ionization corrections to the ionic column densities.  We
estimate the electron density, hydrogen volume density, and the intensity
of ionizing radiation through an analysis of the fine structure
CII*~1335 transition.  Finally, we provide a brief discussion and
summary in $\S$4.

\section{OBSERVATIONS AND ANALYSIS}

We observed Q2231$-$00 ($z_{em} = 3.02$) on the night of 
November 1, 1995 with HIRES (\cite{vgt92}) on the 10m W. M. Keck I Telescope
for a total integration time of 2.5 hours.  We used the C5 decker
plate with $1.1''$ slit, standard $2 \times 1$ binning on the
$2048 \times 2048$ Tektronix CCD, and the kv380 filter. 
For reduction and calibration of the data, we took a 360s
exposure of the standard star BD +28 4211 and images of quartz
and Th-Ar arc lamps.  The data was reduced and wavelength calibrated
with the software package written by Tom Barlow 
and continuum fit with the Iraf package {\it continuum}.

\begin{figure}[ht]
\begin{center}
\includegraphics[height=3.7in, width=2.6in, angle=-90,bb = 55 48 557 744]{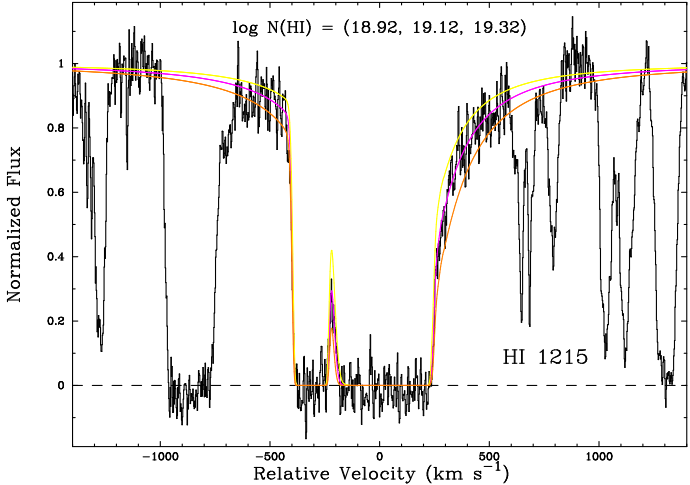}
\caption{The Ly$\alpha$ profile of the Lyman limit system
at $z = 2.652$ toward Q2231$-$00.  The fit was performed with the
VPFIT software package and corresponds to $\log \N{HI} = 19.12 \pm 0.15$~dex
which includes an estimate to the continuum error.}
\label{NHIfig}
\end{center}
\end{figure}

A crucial component in the analysis of any QAL system is the 
neutral hydrogen column density, $\N{HI}$.  Figure~\ref{NHIfig}
shows the HI \lya profile for the LL system at $z = 2.652$
toward Q2231$-$00. 
While the line is heavily saturated, 
the damping wings are resolved and provide a good
measurement of the $\N{HI}$ value, 
$\log \N{HI} = 19.12$~dex.  
The curves in Figure~\ref{NHIfig}
correspond to the line-profile fit (with estimated errors) derived
from the VPFIT software package, kindly provided by R. Carswell and J. Webb.
We include an estimated 0.15~dex error to
the $\N{HI}$ value due to the difficulty in determining 
the unabsorbed continuum flux in the \lya forest.

\begin{figure}[ht]
\begin{center}
\includegraphics[height=5.1in, width=3.7in, bb = 55 48 557 744]{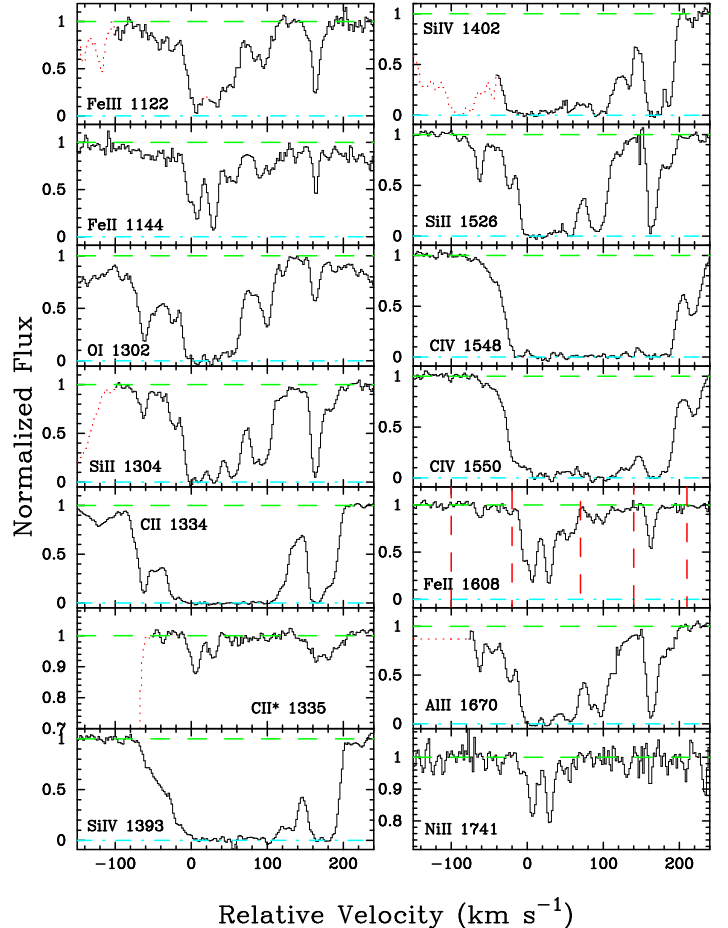}
\caption{Velocity profiles of all the observed
metal-line transitions.  The dotted regions indicate portions
of the profile contaminated by other absorption lines.  The
dashed vertical lines in FeII~1608 define the velocity regions
discussed in the text.}
\label{metalfig}
\end{center}
\end{figure}

Figure~\ref{metalfig} presents the velocity profiles of the metal-line 
transitions observed for this LL system, with $v = 0$ corresponding to 
$z = 2.652$.  The dotted lines designate regions of the
profiles contaminated by other metal lines or \lya forest clouds.
The dashed vertical lines overplotted on the FeII~1608 transition
indicate the velocity regions discussed below (Table~\ref{clm-tab}).
Note that the profiles track one another very closely in velocity space,
{\it even across significantly different ionization states} 
(e.g.\ Fe$^+$, Fe$^{++}$, and Si$^{3+}$).
As emphasized below, this observation is contrary to that typically
observed for the damped \lya systems (\cite{pro96,wol99}).
We have measured the ionic column densities from the unsaturated
and mildly saturated line profiles by summing the apparent column
density 
($N_a(v) = m_e c \ln[I_c/I_a(v)] / \pi e^2 f \lambda$; \cite{sav91})
over several velocity regions.  This approach enables one 
to calculate ionic column densities as a function of velocity to
search for
variations in the ionization state, metallicity, and dust depletion.  
Table~\ref{clm-tab} lists the
column densities for a number of the ions over four velocity regions
and the total integrated column densities.
We are particularly interested in the relative ionic column densities
of Fe$^+$, Fe$^{++}$, Si$^+$, and Si$^{3+}$ because 
the ratios Fe$^+$/Fe$^{++}$ 
and Si$^+$/Si$^{3+}$ constrain the ionization state of the system.

\begin{figure}[hb]
\begin{center}
\includegraphics[height=3.3in, width=3.7in,bb = 55 48 557 744]{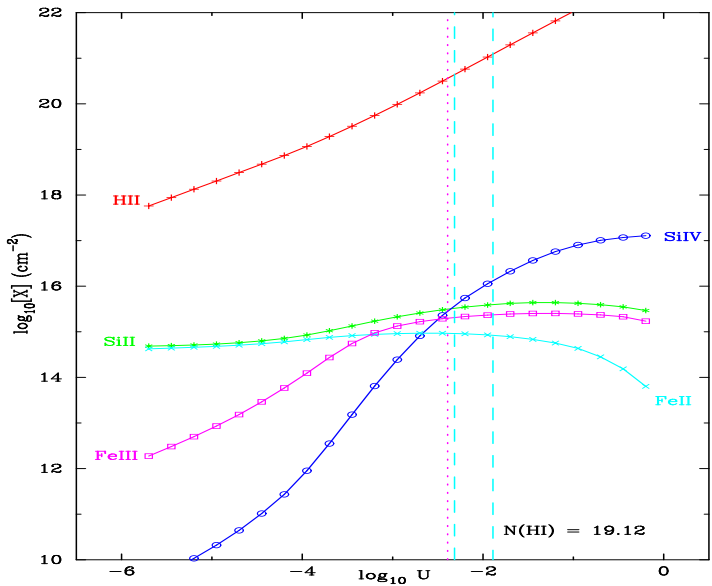}
\caption{The results from a series of CLOUDY calculations (Ferland 1991)
assuming a Haardt-Madau ionizing spectrum and $\log \N{HI} = 19.12$.
Plotted are the calculated mean ionic column densities (assuming intrinsic
solar abundances and [Fe/H] = $-0.5$)
versus a modified ionization parameter, $U$, as defined in the text.
The vertical dashed lines indicate the observed Fe$^+$/Fe$^{++}$ ratio
(with errors) and the dotted vertical line is a lower limit to the
ionization parameter determined from the upper limit to the 
Si$^+$/Si$^{3+}$ ratio.}
\label{cldy}
\end{center}
\end{figure}

To identify the ionization state of this LL system system, we
have utilized the software package CLOUDY v90.04 (\cite{fer91}).
Figure~\ref{cldy} presents the predicted ionic column densities versus
ionization parameter $U$ for a model system assuming a
Haardt-Madau spectrum (\cite{haa96}), where

\begin{equation}
U \equiv {\phi_{912} \over c \, n_H }
= {J_{912} \over 4 \pi h c \, n_H} =
\, (2 \sci{-5}) \; { J_{912} / 10^{-21.5} \over n_H / {\rm cm^{-3}}}
\label{Ueq}
\end{equation}

\noindent with $n_H$ the volume density of Hydrogen,
and $\phi_{912}$  and $J_{912}$ 
the flux and intensity of the incident radiation at 1 Rydberg.
Note this is a modified definition of $U$; it was chosen to
facilitate future comparisons with this work.
The relative ionic column densities were calculated with the CLOUDY package
assuming $\N{HI} = 10^{19.12} \cm{-2}$, intrinsic solar abundances, and
[Fe/H] = $-0.5$~dex.
The dashed vertical lines denote the
range of $U$ values consistent with the Fe$^+$/Fe$^{++}$ ratio
($0.40 \pm 0.05$~dex)
and the dotted line indicates the $U$ value corresponding
to the upper limit on the Si$^+$/Si$^{3+}$ ratio (i.e.\ a lower limit
to $U$). 
These observations imply $\log U = -2.23 \pm .21$, which is 
the most accurate determination of $U$ in a LL system to date.  

\section{PHYSICAL PROPERTIES}

In the previous section, we presented an accurate
measurement of the system's ionization state by comparing the
observed ionic column densities of H$^0$, Fe$^+$, Fe$^{++}$,
Si$^+$, and Si$^{3+}$ with the predicted results from a calculation
performed with the CLOUDY software package.
We find the system is highly ionized with ionization
fraction, $x \equiv {\rm H}^+/{\rm H} = 0.97 \pm 0.02$.  
This implies that the total Hydrogen 
column density, $\log \N{H} = 20.73 \pm 0.2$, 
is comparable to that observed in the damped \lya systems.  
This value is similar to the estimates from Steidel (1990) for his
sample of LL systems.
By making the appropriate ionization corrections
from the CLOUDY results,  
we derive the elemental abundances of this system.
The logarithmic abundance of element X relative to Hydrogen
and normalized to solar abundances (\cite{and89}),
[X/H]~=~${\rm \log [\N{X}/\N{H}] - \log [X/H]_\odot}$, 
are listed in the last column of Table~\ref{clm-tab}.
We find [Fe/H] = $-0.5$~dex which is higher than the typical damped
\lya system and considerably higher than any other estimate to the
abundances of a LL system at $z>2$.

\begin{table*}
\begin{center}
\caption{ \label{clm-tab}}
{\sc Ionic Column Densities and Abundances}
\begin{tabular}{lccccccc}
\tskip \tableline
\tableline \tskip
Ion & $\lambda$ & Reg1\tablenotemark{a} & Reg2\tablenotemark{b} 
& Reg3\tablenotemark{c} & Reg4\tablenotemark{d} & TOT & [X/H] \\
\tableline \tskip
HI & 1215 & ... & ... & ... & $19.12 \pm 0.2$ \nl
CII & 1334 & $14.24 \pm 0.01$ & $15.41 \pm 0.08$\tablenotemark{e} & $>14.76$ 
& $14.42 \pm 0.03$ & $>15.41$ & $> -0.85$ \nl
CII* & 1335 & ... & $12.90 \pm 0.03$ & ... & ... & ... \nl
OI & 1302 & $14.49 \pm 0.02$ & $> 15.20$ & $14.22 \pm 0.02$ & $13.82 \pm 0.02$
& $> 15.33$ & $> -0.6$ \nl
AlII & 1670 & $12.35 \pm 0.01$ & $>13.80$ & $12.88 \pm 0.02$ 
& $12.67 \pm 0.02$ & $> 13.84$ & $> -0.75$ \nl
SiII & 1304 & $13.58 \pm 0.02$ & $>14.87$ & $14.18 \pm 0.01$ 
& $14.03 \pm 0.02$ & $>15.02$ & $> -0.55$ \nl
SiIV & 1402 & ...  & $> 14.50$ & $> 14.21$ & $>14.09$ & $> 14.78$ \nl
FeII & 1608 & $12.96 \pm 0.08$ & $14.34 \pm 0.01$ & $13.38 \pm 0.03$ 
& $13.43 \pm 0.03$ & $14.44 \pm 0.05$ & $-0.5 \pm 0.1$ \nl
FeIII & 1122 & $13.71 \pm 0.04$ & $14.69 \pm 0.03$ & $13.86 \pm 0.03$ 
& $13.83 \pm 0.03$ & $14.84 \pm 0.02$ \nl
NiII & 1741 & ... & ... & ... & ... & $13.36 \pm 0.04$ & $-0.78 \pm 0.05$ \nl
\tskip \tableline
\end{tabular}
\end{center}

\centerline{$^a$Velocity region spanning: $-100 \mkms < v < -20 \mkms$}
\centerline{$^b$Velocity region spanning: $-20 \mkms < v < 70 \mkms$}
\centerline{$^c$Velocity region spanning: $70 \mkms < v < 140 \mkms$}
\centerline{$^d$Velocity region spanning: $140 \mkms < v < 210 \mkms$}
\centerline{$^e$This value was obtained with the method described
in $\S$3}
\end{table*}

To gauge the accuracy with which we
have determined the ionization state and thereby the chemical
abundances,  
we have investigated the effects of varying $\N{HI}$ and the
shape of the input spectrum.
We find that the 0.15~dex uncertainty in the $\N{HI}$ value lends to less than
a 0.1~dex uncertainty in [Fe/H] and the majority of other elemental 
abundances, including the $\N{H}$ value.
Furthermore, we considered the effects of a steeper spectrum
(e.g.\ a Bregman-Harrington spectrum) and find a similar variation in
the measured abundances.
Examining the relative abundances of this system, we observe little
departure from the Solar abundance pattern.  There is, however, evidence for 
an overabundance of Si/Fe (the lower limit to [Si/H] is quite conservative
given the degree of saturation in the Si profiles)
and an underabundance of Ni/Fe as observed 
in the majority of the damped \lya systems.   
In the damped systems,
this pattern is typically interpreted as the result of dust depletion
and/or TypeII supernovae enrichment.
Unfortunately,
the SII~1253 and SII~1259 profiles in this system
are blended with \lya forest
clouds for the [S/Fe] measurement is sensitive to both 
of these interpretations.

The detection of the fine structure CII$^*$~1335 transition allows
further insight into the physical characteristics of this system.
The fine structure line can be excited by several processes, but for
a highly ionized system at this redshift,
electron collisions dominate (\cite{morris86}).  Therefore, the
observed $\N{CII^*}/\N{CII}$ ratio provides a measure of the electron
density via: $\N{CII^*}/\N{CII} = 3.9 \sci{-2} \, n_e \,
\ltk 1 + (0.22 n_p / n_e) \rtk$.  Unfortunately
the $\N{CII^*}/\N{CII}$ ratio cannot be measured directly because the
CII~1334 profile is saturated over the velocity region where 
the CII$^*$ absorption is detected\footnote{We expect the absorption in 
Reg4 ($v \approx 170 \mkms$) for the CII$^*$~1335 transition is due to an
intervening \lya forest cloud and restrict the analysis to the
absorption features in Reg2 ($v \approx 20 \mkms$).}.  We can accurately 
estimate the $\N{CII^*}/\N{CII}$ ratio, however, 
provided the assumption 
that the CII~1334 and FeII~1608 profiles track one another in velocity
space.
Low-ion profiles always track one another in the damped \lya systems
(e.g.\ \cite{pro96,lu96b}) 
and one notes {\it all} of the transitions -- irrespective of ionization
state -- trace one another in this system. 
Our approach, then, is (1) to measure $\N{CII} / \N{FeII}$ 
in Reg4 where the effects of saturation are minimal and  
(2) correct a $\N{FeII}$ measurement in Reg2
by the $\N{CII}/\N{FeII}$ ratio to estimate $\N{CII}$ in this velocity
region.  We find $\N{CII} = 15.41 \pm 0.08 \cm{-2}$
and correspondingly, $\N{CII^*}/\N{CII} = 3.09 \pm 0.75 \sci{-3}$.  
Assuming $n_p/n_e \approx 1$ for this highly ionized system 
this yields an electron density, $n_e = 6.5 \pm 1.3 \sci{-2} \cm{-3}$.

The $n_e$ value coupled with our knowledge of the ionization
state enables an estimate of the hydrogen volume density and the
intensity of the ionizing radiation.  Ignoring the contribution
to $n_e$ from metals (a reasonable assumption when $x \approx 1$)
and adopting $\N{He^+}/\N{He} = 0.9$ and $\N{He^{++}} / \N{He} = 0.1$
from the CLOUDY calculations, we find 
$n_H = 5.9 \pm 1.2 \sci{-2} \cm{-3}$.  With $\N{H} \approx 10^{20.73} \cm{-2}$,
this implies a path length $\ell \equiv \N{H}/n_H \approx 3 \pm 1.6$~kpc.  
While this length is rather uncertain ($> 50\%$ statistical error) and
does not account for clumping along the sightline, it is still one
of the most meaningful size estimates of a QAL system to date.
The ionizing intensity,
$J_{912}$, is easily determined by inverting Equation~\ref{Ueq}:
$\log J_{912} = -20.22 \pm 0.21$ which is higher (a factor of $5-10$)
than typical estimates
from the Proximity Effect (\cite{lu91}), but not unreasonably high.
Perhaps there is a source of local ionizing radiation 
(e.g.\ supernovae, OB stars) in addition to the extragalactic background
radiation.
Finally, consider the kinematic characteristics of this LL system.
First note that the width of the profile spans nearly 200~\kms.
Second, we observe that the strongest absorption feature lies at the
left edge of the profile and the optical depth decreases 
monotonically towards positive velocity.  This edge-leading asymmetric
trait is characteristic of that observed in the damped \lya systems
(\cite{pro97b,pro98}).  Contrary to the damped \lya systems, however,
the low-ion profiles trace very closely the high-ion profiles
(Si$^{3+}$ and Fe$^{++}$).  
This indicates the various ionization states arise from the same
regions within the system.  Interestingly, one observes a similar
correspondence between the line profiles of
multiple ionization states in the 
ISM (\cite{sav97}).
A future survey will determine
if a large sample of LL systems exhibit similar traits.

\section{SUMMARY AND CONCLUSIONS}

We have presented techniques for accurately determining the ionization
state of a LL system.  The technique
requires the measurement of multiple ionization stages of a single element --
in this case Fe$^+$/Fe$^{++}$ -- an estimate of $\N{HI}$ 
and a series of calculations with the CLOUDY software package.  
It should be applicable to a number of high redshift 
LL systems and will enable a survey of their elemental abundances.
For this system, we find an ionization fraction
$x = 0.97 \pm 0.02$ and a corresponding Iron abundance,
[Fe/H] = $-0.5 \pm 0.1$~dex.   The overall abundance pattern resembles
that observed in the damped \lya system possibly suggesting an underlying
depletion pattern or TypeII SN enrichment.
This system is unusual for exhibiting CII*~1335 absorption which  
provides one with an estimate of the electron density
($n_e = 6.5 \sci{-2} \cm{-3}$).  Our knowledge of the ionization state
allows estimates of the hydrogen volume density 
($n_H = 5.9 \sci{-2} \cm{-3}$), the intensity of the ionizing
radiation ($\log J_{912} = -20.22$), and the path length through the
system $\ell = 3 \pm 1.6$~kpc.

It is instructive to compare this LL system with the properties of the
damped \lya systems.  For example, we note
the derived $\N{H}$ and $n_H$ values
are similar to those observed for the damped systems.  In fact,
if it were not for the high intensity of ionizing radiation, 
this system would be primarily neutral with $\N{HI} > 2 \sci{20} \cm{-2}$.
The observed abundances and kinematic characteristics
further argue for this interpretation.  
It will be exciting to investigate the connection between other LL
systems and the damped \lya systems
through a survey of LL systems at $z>2$.  This research will help develop
a better understanding of the connection between the QAL systems
and galaxy formation. 
In addition, a direct comparison with numerical simulations 
may provide insight into the nature of protogalaxies in the early 
universe.

\acknowledgements

We wish to acknowledge 
A. Wolfe, S. Burles and F. Hamann for helpful comments.
We would also like to thank Bob Carswell and John Webb for
kindly providing the VPFIT package, Piero Madau for providing an
electronic version of their UVB spectrum and G. Ferland for 
developing and distributing CLOUDY. 
We thank T. Barlow for developing and providing the HIRES
data reduction package.
Finally, this work could not have been carried out without the helpful
support of the Keck observing staff.
JXP was partially supported by 
NASA grant NAGW-2119 and NSF grant AST 86-9420443.

\end{document}